\documentclass[a4paper,11pt]{article}
\pdfoutput=0
\usepackage{jcappub} % for details on the use of the package, please
\usepackage[T1]{fontenc}
\def\be{\begin{equation}}
\def\ee{\end{equation}}
\def\bea{\begin{array}}
\def\eea{\end{array}}
\def\beqa{\begin{eqnarray}}
\def\eeqa{\end{eqnarray}}
\def\beqas{\begin{eqnarray*}}
\def\eeqas{\end{eqnarray*}}

\def\bp{\begin{picture}}
\def\ep{\end{picture}}
\def\bc{\begin{center}}
\def\ec{\end{center}}
\def\bfig{\begin{figure}}
\def\efig{\end{figure}}

\def\bit{\begin{itemize}}
\def\eit{\end{itemize}}
\def\nn{\nonumber}
\def\f{\frac}

\def\[{\left[}
\def\]{\right]}
\def\({\left(}
\def\){\right)}

\def\..{\left.}
\def\.{\right.}
\def\tl{\tilde}
\def\ra{\rightarrow}

\def\tm{\times}

\def\da{\dagger}

\def\ep{\epsilon}

\def\pa{\partial}
\def\pr{\prime}

\newcommand{\vecs}[1]{\mbox{\boldmath${#1}$}}

\title{Q-balls Formation and the Production of Gravitational Waves With Non-minimal Gravitational Coupling}
\author[]{Fei Wang$^1$,\note{Corresponding author.}}
\author[]{Rui Wang}
\affiliation[]{Department of Physics, Zhengzhou University, No. 100 Science Avenue,\\ ZhengZhou 450001, P.R.China}
\emailAdd{feiwang@zzu.edu.cn}
\emailAdd{rwang@stu.xmu.edu.cn}
\abstract{We propose to introduce non-minimal couplings of Affleck-Dine (AD) field to gravity by adding the
coupling of AD field to the Ricci scalar curvature.
As the Jordan frame supergravity always predict $|\Phi|^2 {\cal R}/6$ type coupling for scalars with canonical kinetic terms, we propose a way to realize the required $c_0|\Phi|^2 {\cal R}$-type couplings with generic $c_0$ for canonical complex scalar fields  after SUSY breaking.
The impacts of such non-minimal gravitational couplings for AD field is shown, especially
on the Q-balls formation and the associated gravitational wave (GW) productions.
New form of scalar potential for AD field in the Einstein frame is obtained.
By numerical simulations, we find that, with non-minimal gravitational coupling to AD field,
Q-balls can successfully form even with the choice of non-negative $K$ parameter for $\xi>0$.
The associated GW productions as well as their dependences on the $\xi$ parameter are also discussed.}
\begin{document}
\maketitle
\flushbottom
\section{Introduction}
It is well known that the baryons makes up 5 percent of the total energy density of the universe with little or no primordial antimatter.
To explain the origin of baryon asymmetry, Sakharov suggested that the baryon asymmetry might be understandable in terms of microphysical laws instead of some sort of initial condition.
 A small baryon asymmetry can be produced in the early universe if three necessary conditions are satisfied: the baryon number violation, C and CP violation,
the departure from thermal equilibrium~\cite{sakharov}. Many baryogenesis mechanisms had been proposed so far,
for example, the electroweak baryogenesis~\cite{EWBAU}, the leptogenesis~\cite{leptogenesis},
the GUT baryogenesis~\cite{BAU:GUT} and the Affleck-Dine (AD) mechanism of baryogenesis~\cite{AD:mechanism} (See~\cite{BAU:review1,BAU:review2} for excellent reviews). To understand better those baryogenesis mechanisms,
it is interesting to survey their various cosmological consequences and observation signals, for example, their possible gravitational wave (GW) signals.

Observations of the GWs from black hole mergers by the Advanced LIGO/VIGO detectors~\cite{LIGO} open a new era in astrophysics and cosmology. Many experiments~\cite{LISA,DECIGO} plan to further explore GWs, including both the transient GW signals and stochastic GW backgrounds,
in a broader range of frequencies and with more accuracy in the coming decades.
The stochastic GW backgrounds could reveal certain interesting properties  of the very early universe, including the information of the baryogenesis stage, because the relevant dynamics can be potential sources of the stochastic GW backgrounds. Different types of phase transitions other than  the electroweak and the QCD phase transitions can also possibly appear in our cosmic history. Their
applications and experimental signatures, in particular in the context of exciting GWs, are discussed in~\cite{Mazumdar:2018dfl}, which could potentially be constrained by LIGO/VIRGO, Kagra, and eLISA.

%%In AD mechanism, the coherent production of baryons or leptons by the AD  fields~\cite{AD:mechanism} can satisfy Sakharov's conditions.
 In AD mechanism of baryogenesis, the AD field starts oscillating around its origin and gives rise to rotational motion when the Hubble parameter becomes as small as the AD scalar mass after inflation, making the baryon number of the universe.
The instability of AD field oscillations under small perturbations, which are inevitably
introduced by quantum fluctuations of the field, will drive the condensate to fragment into
non-topological solitons called Q-balls~\cite{Qballs}.
Most of the baryon asymmetry is absorbed into Q-balls. The existence and stability of such non-topological solitons are guaranteed by the conserved charge related to a global symmetry.
Many numerical studies simulate the formation of Q-balls~\cite{simulation:Qball,Q-ball:gauge,Q-ball:gravity,Q-ball:other}.
Q-balls can act as good dark matter candidate and give a possible explanation on the coincidence
problem of baryonic matter and dark matter relic abundances~\cite{Q-ball:coincidence1,Q-ball:coincidence2,Q-ball:coincidence3,Q-ball:coincidence4}.

The formation of AD condensate is fairly generic, relying only on the assumptions of inflation and flat directions~\cite{flat:direction}, which naturally appear in the context of supersymmetry (SUSY). SUSY is widely regarded as one of the most promising candidates for new physics beyond the standard model. A generic feature of SUSY gauge theories is that the size of the VEVs of scalar fields are not fixed, but usually have some extended range. There are a number of directions in the
space of scalar fields, collectively called the flat directions, where the scalar potential vanishes identically in the global SUSY limit except for possible nonrenormalizable operators in the
superpotential~\cite{fd:lift}. Flat directions can be parameterized by some gauge invariant combinations of squarks and sleptons. Such flat directions can act as the AD fields for baryogenesis.

It was shown~\cite{Q-ball:GW1,Q-ball:GW2,Q-ball:GW3} that significant GWs can be emitted during the Q-balls formation associated with AD mechanism of baryogenesis because the formation of Q-balls is inhomogeneous and not spherical. The AD condentate can also generate isocurvature fluctuations~\cite{Mazumdar:2010sa}, which also potentially emit GWs. The formation of Q-balls depend on the effective potential of the AD field, which had been discussed for different SUSY breaking
types~\cite{Q-ball:gauge,Q-ball:gravity,Q-ball:other}. On the other hand, the previous
discussions on Q-balls formation are carried out in the Einstein frame. It is interesting to
check the Q-ball formation processes in the Jordan frame, in which the AD field can have a direct
coupling to the Ricci scalar $R$. Transforming from the Jordan frame to the Einstein frame, the shape
of the effective scalar potential for AD field can be changed and the GW signals can show a different
pattern. So, it is interesting to discuss the impact of such non-minimal couplings of AD fields to
gravity in the AD mechanism and know the new predictions of this scenario.
In AD inflation~\cite{AD:inflation}, similar non-minimal coupings of AD field to gravity had also
been used. Besides, GWs emitted during the Q-balls formation can carry important informations related
to the interactions of AD field. So, if such GW signals are observed, they will provide a new tool for
our understanding of AD baryogenesis mechanism.

%%The AD inflation are also discussed in 1909.12300 and 2011.10397.
This paper is organized as follows. In Sec \ref{sec-2},
we discuss the consistency of the non-minimal coupling terms in Jordan frame supergravity after SUSY breaking.
Then we deduce the form of the scalar potential for AD fields in Einstein frame.
In Sec \ref{sec-3}, we discuss and simulate numerically the formation of Q-balls and the associated GW
productions. Sec \ref{sec-4} contains our conclusions.

\section{Affleck-Dine fields with non-minimal gravitational coupling}
\label{sec-2}

We want to introduce the non-minimal couplings of AD field to gravity. A coupling of AD field to the Ricci scalar curvature ${\cal R}$ can be present in Lagrangian,
 just as the case in the Higgs inflation, in which a large non-minimal coupling of
the Higgs doublet to gravity is introduced (in the Jordan frame). In fact, in non-SUSY version, such a term can always emerge from the graviton-scalar loops if we adopt the effective field theory treatment of quantum gravity. However, in SUSY extension, such a $c_0|\Phi|^2 {\cal R}$-type coupling with arbitary $c_0$ is difficult to generate, which always give $c_0=1/6$ for canonical scalar kinetic terms~\cite{1004.0712}. In the SUSY framework, as far as we know, no discussions on realization of such a general form had been given in the literatures. We propose to generate such a form with typical SUSY breaking terms.
\subsection{Generating the $c_0|\Phi|^2 {\cal R}$-type coupling after SUSY breaking}

In the SUSY version of non-minimal coupling of scalar to gravity, a Jordan frame scalar-gravity
action needs to be supersymmetrized. The natural starting point is not global supersymmetry but the modification of the Lagrangian for supergravity coupled
to a multiplet of chiral superfields in the Jordan frame. The general recipe for the formulation of 4D Jordan frame supergravity was discussed
in~\cite{1004.0712,1005.2735,1008.2942,1203.0805}. A complete explicit $N=1$, $d=4$ supergravity action in an arbitrary
Jordan frame with non-minimal scalar-curvature coupling of the form $\tl{\Phi}(z, \bar z)\, {\cal R}$
had been constructed in ~\cite{1004.0712}. The bosonic part of the Jordan frame supergravity
Lagrangian is
 \begin{equation}
\mathcal{L}_{J}^{bos,\mathcal{K}(\tl{\Phi} )}=\sqrt{-g_{J}}\left[ -{\textstyle
\frac{1}{6}}\tl{\Phi} {\cal R}(g_{J})-\tl{\Phi}_{\alpha\bar\beta }\hat{\partial}
_\mu z^\alpha\hat{\partial}^\mu \bar z^{\bar\beta }+\tl{\Phi}
\mathcal{A}_\mu^2-V_{J}+\mathcal{L}_{1}^{bos,
\mathcal{K}(\tl{\Phi})}\right] \,,  \label{Jordan:SUGRA}
\end{equation}
with the frame function $\tl{\Phi}(z, \bar z)$ related to the K{\"a}hler potential $\mathcal{K}(z,\bar z)$ by
\begin{equation}
\mathcal{K}(\tl{\Phi}(z,{\bar z}))= -3\log (-\f{1}{3}\tl{\Phi}(z,{\bar z}))\,.
\label{K}
\end{equation}
As noted there, in order to have canonical kinetic terms in the Jordan frame, it is sufficient to
take the following form of the frame function
\beqa
\tl{\Phi}(z,{\bar z}) =-3 + \delta_{\alpha\bar\beta }z^{\alpha} \bar z^{\bar\beta } +J(z) +\bar J({\bar z})\,,
\eeqa
and requires that the bosonic part of the auxiliary vector field to vanish
\begin{equation}
\mathcal{A}_\mu =0\,.
\end{equation}
To break the superconformal symmetry of the matter multiplets in the Jordan frame
supergravity action (without introducing dimensional parameters into the underlying superconformal
action), which take the conformal coupled form $\f{1}{6} |\Phi|^2 {\cal R}$ in the scalar-gravity part,
one can modify the real function frame function with additional holomorphic $J_{\cal O}$ (anti-holomporhic $\bar{J}_{\cal O}$)
function terms. However, with the modified frame function, the scalar-gravity Lagrangian will
have the additional term $c_0\(J_{\cal O}|_{s}+ \bar{J}_{\cal O}|_{s}\){\cal R}$ in addition to
$\f{1}{6} |\Phi|^2 {\cal R}$.  For scalar bilinear $J_{\cal O}|_{s}$, such new non-minimal coupling
contribution is not of the $c_0|\Phi|^2 {\cal R}$ form. However, as far as we know, no discussions on the realization of such a form in the SUSY framework had been given in the literatures. We note that, after SUSY breaking, coupling of the form $c_0|\Phi|^2 {\cal R}$ could be generated consistently.
%%%The ordinary $c_0|\Phi|^2 R$ form can be generated in the low energy Jordan frame scalar-gravity Lagrangian through proper SUSY breaking terms.
%% We can illustrate this with a toy model. Consider a U(1) %% form will generate the coupling $|\Phi|^2 J_{\cal O}|_s$, which can generate the effective $|\Phi|^2 R$ term at loop level or after integrating
%%%%out the heavy superfields within $J_{\cal O}$.

We propose a way to construct a $\xi |\Phi|^2 {\cal R}$-type term in the scalar-gravity part of
Jordan frame supergravity after SUSY breaking.
Assume that there are two chiral superfields $P,\tl{P}$, which can take opposite gauge quantum numbers,
for example, the fundamental and antifundamental representation of $SU(N)$, respectively.
We can add holomorphic (and anti-holomorphic) terms in the Kahler potential of the two chiral fields, which is given as
\beqa
K\supseteq {P}^\da P+\tl{P}^\da \tl{P}+ \(c_P \tl{P} P+ h.c.\)~,
\label{Kahler:SUSYB}
\eeqa
or a bilinear $\mu\tl{P}P$ term in the superpotential.
After SUSY breaking, the soft SUSY breaking scalar bilinear B-term $B_0\mu \tl{P}P$ can be generated.
For example, in the anomaly mediation SUSY breaking mechanism,
after rescaling
\beqa P\ra \phi P~,~~~\tl{P}\ra \phi\tl{P}~,\nn\eeqa
with $\phi=1+\theta^2 F_{\phi}$ the compensator field and $F_\phi\simeq m_{3/2}$, a $B\mu$-term $c_P m_{3/2}^2\mu \tl{P}P$ will be generated from eqn.(\ref{Kahler:SUSYB}) in addition to a $c_P m_{3/2} \tl{P} P$ term in the superpotential~\cite{KahlerD:NW,KahlerD:luty,KahlerD:Fei}.
The scalar part of the chiral superfields is givens as
\beqa
{\cal L}\supseteq (\pa_\mu P_s)^\da (\pa^\mu P_s)+(\pa_\mu \tl{P}_s)^\da (\pa^\mu \tl{P}_s)
-|c_P m_{3/2}|^2\(|P_s|^2+|\tl{P}_s|^2\)-\(c_P |m_{3/2}|^2 \tl{P}_sP_s+h.c.\),~\nn\\
\eeqa
with $P_s$ and $\tl{P}_s$ being the scalar components of superfields $P$ and $\tl{P}$, respectively.
From the previous expressions, we can get the scalar mass matrix
\beqa
\(P_s^\da~,~~\tl{P}_s\)\(\bea{cc}~c^2_P m^2_{3/2}~&~c_P m^2_{3/2}~\\~c_P m^2_{3/2}~&~c_P^2 m^2_{3/2}~\eea\)\(\bea{c}P_s\\\tl{P}_s^\da\eea\)~.
\eeqa
Here ${P}_s,\tl{P}_s$
denote the scalar part of superfields $P,{P}^\pr$, respectively.
 We can redefine new scalar fields as
\beqa
\phi_1&=&\f{1}{\sqrt{2}}\( {P}_s+\tl{P}^\da_s \)~,~~~~
\phi_2=\f{1}{\sqrt{2}}\({P}_s-\tl{P}^\da_s\)~,\nn\\
\phi_1^\da&=&\f{1}{\sqrt{2}}\( {P}^\da_s+\tl{P}_s \)~,~~~~
\phi_2^\da=\f{1}{\sqrt{2}}\({P}_s^\da-\tl{P}_s\)~,
\eeqa
so that the scalar mass mixing terms are removed. The corresponding eigenvalues for the scalars $\phi_1$ and $\phi_2$ are $(c_P^2\mp c_P) m_{3/2}^2$.

The framefunction for $P$ and $\tl{P}$, with modification by bilinear holomorphic $J_{\cal O}$ term and antiholomorphic $J_{\cal O}^\da$ terms,
are given as
\beqa
\tl{\Phi}(P,{P}^\da;\tl{P},\tl{P}^\da)&\simeq& -3e^{- K(P,{P}^\da;\tl{P},\tl{P}^\da)/3}\nn\\
&=& -3 + P^\da P+\tl{P}^\da \tl{P} +\(c_P \tl{P} P+ h.c.\)~.
\eeqa
After SUSY breaking via anomaly mediation (as discussed previously), the corresponding Kahler potential will lead to canonical kinetic terms for scalar components of $P$ and $\tl{P}$, consequently also the new $\phi_1,\phi_2$ scalar fields. The Jordan frame scalar-gravity coupling $-\tl{\Phi}(P_s,{P}^\da_s;\tl{P}_s,\tl{P}_s^\da)\cdot {\cal R}(g_J)/6$
can be rewritten in terms of the new canonical $\phi_1,\phi_2,{\phi}^\da_1,{\phi}^\da_2$ variables
\beqa
&&-\f{{\cal R}(g_J)}{6}\tl{\Phi}(\phi_1,\phi_2,{\phi}^\da_1,{\phi}^\da_2)\nn\\
&=&-\f{{\cal R}(g_J)}{6}\left[-3+\(1-c_P\)\({\phi}^\da_1{\phi}_1\)+
\(1+c_P\)\({\phi}^\da_2{\phi}_2\)\right].
\eeqa
 It can be seen that the general $\xi |\Phi|^2 {\cal R}$-type form with $\xi\neq 1/6$ can be obtained with proper choices of $c_P$ parameter after SUSY breaking.
%%, assuming one of the canonical scalar is stabilized at $\phi_a=0$.

\subsection{Scalar potential with non-minial gravitational coupling of AD field}
The scalar-gravity part in the Jordan frame
can be written as
\beqa
\label{Jordan}
S \supseteq \int d^4x \sqrt{-g}\[\f{M_P^2}{2}{\cal R}+ (\pa^\mu \Phi^*) (\pa_\mu \Phi) +\xi |\Phi|^2{\cal  R} -V(|\Phi|)\]~,
\eeqa
with $V(|\Phi|)$ the scalar potential, taking into account the SUSY breaking effects.
For example, in gravity-mediated SUSY breaking scenarios, the scalar potential for AD field
takes the form~\cite{Q-ball:gravity}
\begin{equation}
    \label{potential:AD}
    V(|\Phi|) = m_{3/2}^2 |\Phi|^2
    \left[ 1 + K\log\left(\frac{|\Phi|^2}{M_P^2}\right)\right ]+ Am_{3/2}\left(\frac{\Phi^{d}}{dM_P^{d-3}}+h.c.\right)
    +\frac{|\Phi|^{2d-2}}{M_P^{2d-6}} \,.
\end{equation}
The value of $K$ in eq.(\ref{potential:AD}) can be computed as
 \be\label{K1}
K=\frac{1}{q^2}{\partial m_S^2\over \partial t}\Big\vert_{t={\log} q}~,
\ee
with
\beqa
m_S^2 = \sum_{a} p_i^2 m_i^2~,~~~\sum p_i^2=1~.
\eeqa
From the renormalization group equation~(RGE) of soft SUSY breaking scalar masses, the beta functions can be seen to take the following form at the one-loop level
\be\label{rge}
{\partial m_i^2\over \partial t}=\sum_{g}a_{ig} m_g^2+
\sum_a h^2_a(\sum_j b_{ij}m_j^2+A^2)~.
\ee
Depending on the RGE of $m_i^2$, the sign of the parameter $K$ can be positive if the top quark loop effects are dominant, which is realized when the top Yukawa coupling is order unity. On the other hand, $K$ is negative when the gaugino loop effects are dominant~\cite{sign:K}. AD field with positive $K$ can play an important role in our following discussions.

To eliminate the $\xi |\Phi|^2 {\cal R}$ term in (\ref{Jordan}), we need to make a proper Weyl transformation
to change from the Jordan frame to the Einstein frame. Such a technique always appear in Higgs-inflation type models~\cite{Higgs:inflation}.
We rewrite the complex scalar field $\Phi$ as $\Phi=R e^{i\Theta}/\sqrt{2}$ to manifest the $U(1)$ global symmetry and simplify the following deductions.
The following Weyl transformation is adopted
\beqa
g_{\mu\nu}\ra \Omega^2 g_{\mu\nu},~~~~~~
\Omega^{2}\equiv1+\xi\f{R^2}{M_{P}^2} ~,
\eeqa
to change from the Jordan frame to the Einstein frame.
The kinetic term of $\Phi$ will also change according to the Weyl transformations. As noted in ~\cite{GarciaBellido:2008ab}, a new kinetic term will be generated
\beqa
\f{M_P^2}{2}\[-\f{3}{2}\tl{g}^{\mu\nu}\pa_\mu \ln\Omega^2\pa_\nu \ln\Omega^2\]
&=&-3\f{1}{\Omega^4}\f{\xi^2}{2M_P^2}\f{1}{2}\[4R^2(\pa R)^2\]~,
\eeqa
in addition to the rescaled standard kinetic term in the $R,\Theta$ variable
\beqa
&&{\cal L}_{kinetic}\supseteq -\f{1}{2\Omega^2}\(\pa  R \pa R+ R^2 \pa \Theta  \pa \Theta\).
\eeqa
We need to rescale the $R$ mode to obtain the canonical kinetic expressions.
We can define the new normalized field $\tl{R},\tl{\Theta}$ with the canonical kinetic term
 \beqa
 {\cal L}_{kinetic}\supseteq -\f{1}{2}\(\pa  \tl{R} \pa \tl{R}+\tl{R}^2\pa \tl{\Theta}  \pa \tl{\Theta}\)~,
 \eeqa
 by solving the differential equations
\beqa
\f{d \tl{R} }{d R}&=&\sqrt{\f{\Omega^2+6\xi^2R^2/M_{P}^2}{\Omega^4}}~,
~~~~~\f{d\tl{\Theta}}{d\Theta}=\f{R}{\Omega \tl{R}}~.
\eeqa
The differential equation for $\tl{R}$ can be solved to give
\beqa
\tl{R}&=&F(\xi) -\frac{M_P}{\sqrt{\xi}}\left(\sqrt{6\xi} \tan^{-1}\left[\frac{\sqrt{6} R \xi}{\sqrt{M_P^2+R^2\xi(1+6 \xi)}}\right]\right.\nn\\
&&-\left.\sqrt{1+6 {\xi}} \ln\left[{R} {\xi} (1+6 {\xi})+\sqrt{\xi(1+6 {\xi})} \sqrt{M_P^2+R^2 \xi (1+6\xi)}\right]\right),
\eeqa
with $F(\xi)$ a function of $\xi$ only. We need to choose the form of $F(\xi)$ to be
\beqa
F(\xi)=-\f{M_P\(\ln M_P+\ln \sqrt{\xi}\)+\sqrt{1+6\xi} \ln\( M_P\sqrt{(1+6\xi)\xi}\)}{\sqrt{\xi}},
\eeqa so as that $\tl{R}$ will tend to $R$ both in the $\xi\ra 0$ and $R\ra 0$ limit.
%%%The solution can then be expanded into $\xi$ series
%\beqa\tl{R}=R-\f{R^3}{6M_P^2}\xi+\left(\f{R^3}{M_P^2}+\frac{3 R^5}{40 M_P^4}\right) {\xi}^2+O[\xi]^{5/2}.\eeqa

For $\xi\sim {\cal O}(1)$, we can approximately solve for $\tl{R}$
\beqa
R &\approx& \tl{R}-\f{6\xi^2-\xi}{6M_P^2}\tl{R}^3+\cdots~.
\label{transf:s}\eeqa
The expressions for $\Theta$, in terms of $\tl{\Theta}$ and $\tl{R}$ variables, can be given approximately by
\beqa
{\Theta}\approx \tl{\Theta}+\f{6\xi^2+5\xi}{6M_P^2}\tl{R}^2 \tl{\Theta}.\label{transf:t}
\eeqa
So we have
\beqa
&&\Phi^n+h.c.=\f{R^n}{2^{n/2}}\cos(n\Theta)~\nn\\
&\approx& \f{\tl{R}^n}{2^{n/2-1}}\[1-n\f{6\xi^2-\xi}{6M_P^2}\tl{R}^2\] \[\cos (n\tl{\Theta})- n\f{6\xi^2+5\xi}{6M_P^2}\tl{R}^2 \tl{\Theta}\sin( n\tl{\Theta})\].
\eeqa

The scalar potential for AD fields in the Einstein frame will take the form
\beqa
V^{\pr}(\tl{R},\tl{\Theta})=\f{1}{\Omega^4}V(R(\tl{R}),\Theta(\tl{\Theta},\tl{R})),
\label{xi:potential}
\eeqa
after the substitution of eq.(\ref{transf:s}), eq.(\ref{transf:t}) and the form of $\Omega^2$ in $\tl{\Theta},\tl{R}$ variables
\beqa
\Omega^2=1+\xi\f{{R}^2}{M_{P}^2}\approx 1+\xi\f{\tl{R}^2}{M_{P}^2}-\xi\f{6\xi^2-\xi}{3M_P^4}\tl{R}^4.
\eeqa

\section{Q-balls formation with non-minimal gravitational couplings and GWs}\label{sec-3}

 The AD field can develop a large VEV during inflation, and it starts to oscillate
after inflation when the cosmic expansion rate becomes comparable to its mass.
Soon after the onset of oscillations, the AD field experiences spatial instabilities and deforms into clumpy Q-balls. From the equations of motion for the homogeneous modes and the fluctuations,
one can check if the fluctuations can grow exponentially so as to go nonlinear
and eventually form Q-balls,
given the explicit form of the AD scalar potential.
\subsection{General discussions on the effects of the non-minimal gravitational couplings}\label{3.1}
In the case with the soft SUSY breaking effects from gravity-mediation,
the formation of Q-balls can be possible only with negative $K$ .
This amounts to a potential of $|\Phi|^{2+K}$ shape (with $|K|\ll1$) that is shallower than the quadratic shape. With the presence of non-minimal couplings to Ricci scalar for $\xi\neq 0$, new possibilities
can emerge. For $\xi\sim {\cal O}(1)$ and $\tl{\phi}_i/M_P\ll 1$, the scalar potential for AD fields in the Einstein frame can be
approximately expanded as
\beqa
V^{\pr}(\tl{\phi}_1,\tl{\phi}_2)&\approx&
(1-2\xi\f{\tl{\phi}_1^2+\tl{\phi}_2^2}{M_{P}^2}+\cdots)
V(\tl{R}(\tl{\phi}_1,\tl{\phi}_2),\tl{\Theta}(\tl{\phi}_1,\tl{\phi}_2))~,\\
&\approx& V(\tl{R}(\tl{\phi}_1,\tl{\phi}_2),\tl{\Theta}(\tl{\phi}_1,\tl{\phi}_2))
-2\xi\f{\tl{\phi}_1^2+\tl{\phi}_2^2}{M_{P}^2}V(\tl{R}(\tl{\phi}_1,\tl{\phi}_2),\tl{\Theta}(\tl{\phi}_1,\tl{\phi}_2))+\cdots. \nn
\label{xi:potential2}
\eeqa
Here we rewrite the potential from $\tl{R},\tl{\Theta}$ variables of the new canonical fields into its real and imaginary components $\tl{\phi}_1,\tl{\phi}_2$ for convenience with $\tl{R}^2=\tl{\phi}_1^2+\tl{\phi}_2^2$.
It is obvious that additional suppressed $\tl{\phi}_i^4,\tl{\phi}_i^6,\cdots$ type terms will appear in
the new scalar potential. The emergence of such suppressed $\tl{\phi}_i^n$ terms, with a negative
coefficient for the leading suppressed $\tl{\phi}^4$ term in the case of $\xi>0$,
can possibly make the scalar potential shallower than quadratic form so as that $V(\Phi)/|\Phi|^2$
 has a minimum at $|\Phi|\neq 0$. Therefore, we can anticipate that Q-balls can possibly
form in the case of $K=0$ for positive $\xi$, even in the case with positive $K$ for positive $\xi$.
Numerical results in fact confirm this observation.

 We have the following discussions:
\bit
\item $K<0$ with $\xi>0$.
Similar evolution behavior to the case without non-minimal couplings to Ricci scalar
can be obtained. The additional (leading) suppressed negative $\tl{\phi}_i^4$ term can make the Q-balls formation easier.
\item $K<0$ with $\xi<0$.
 The additional suppressed
positive $\tl{\phi}_i^4$ term will make the Q-balls formation a bit harder.
We need a more negative $K$ to guarantee the formation of Q-balls.
\item $K\geq 0$ with $\xi<0$.
The scalar potential can not be shallower than quadratic type. So, Q-balls can not form in this case.
\item $K\geq 0$ with $\xi>0$.
It is well known that Q-balls are unable to form in the case $\xi=0$ with non-negative $K$.
However, with non-minimal gravitational couplings of AD field to Ricci scalar, numerical results
indicate that Q-balls formation still allows non-negative value of $K$.
It is a unique new feature of our scenario with non-minimal gravitational couplings to AD field.
\eit

\subsection{Gravitational Waves From Q-Ball Formation}\label{3.2}
The fragmentation of the AD field is inhomogeneous and non-spherical, so GWs will be emitted in the process of the Q-balls formation. We will discuss the generation of the GWs associated with the fragmentation of the AD field with non-minimal gravitational coupling and
estimate the amplitudes, the frequencies of the GWs.

In a spatially-flat FRW background, gravitational waves may be represented by the transverse-traceless(TT) part of the spatial gauge-invariant metric perturbation
\beqa
ds^2=-dt^2+a^2(t)(\delta_{ij}+ h_{ij}) dx^i d x^j~,
\eeqa
with $a(t)$ the scale factor and the tensor perturbations satisfying $\pa_i h_{ij}=h_{ii}=0$.
  The initial amplitudes and frequencies of the GWs from the Q-ball formation can be evaluated
  by using the equations for the TT component of the metric perturbations.
 We follow the discussions given in~\cite{GW:juan,Q-ball:GW2,chiba}.
  The perturbed Einstein equations describe the evolution of the tensor perturbations as
\begin{equation}
{\ddot h_{ij}}({\vecs x},t)+3H{\dot h_{ij}}({\vecs x},t)-\frac{\nabla^2}{a^2}h_{ij}({\vecs x},t)
=16 \pi G \Pi_{ij}({\vecs x},t),  \label{evo1}
\end{equation}
where $h_{ij}({\vecs x},t)$ is the TT component of the metric
perturbation and $\Pi_{ij}({\vecs x},t)$ is the TT component of the
energy-momentum tensor of the AD field. It is easier to use the following equations for $u_{ij}$, whose source is the
complete $T_{ij}$ instead of its TT part $\Pi_{ij}$
\begin{equation}
{\ddot u_{ij}}({\vecs k},t)+3H{\dot u_{ij}}({\vecs k},t)+\frac{{\vecs k}^2}{a^2}u_{ij}({\vecs k},t) =16 \pi G T_{ij}({\vecs k},t),   \label{evo2}
\end{equation}
with the relations of $u_{ij}({\vecs k},t)$ and $T_{ij}({\vecs k},t)$
\begin{align}
h_{ij}({\vecs k},t) &= \Lambda_{ij,mn}(\hat{\vecs k}) u_{mn}({\vecs k},t), \\
\Pi_{ij}({\vecs k},t) &=\Lambda_{ij,mn}(\hat{\vecs k}) T_{mn}({\vecs k},t), \\
T_{ij}({\vecs x},t)& = \frac{1}{a^2} \nabla_i \phi({\vecs x},t) \nabla_j \phi({\vecs x},t),
\end{align}
and the projection tensor $\Lambda_{ij,mn}$ defined by
\begin{align}
\Lambda_{ij,mn}(\hat{\vecs k})\equiv \left(P_{im}(\hat{\vecs k})P_{jn}(\hat{\vecs k})-\frac{1}{2}P_{ij}(\hat{\vecs k})P_{mn}(\hat{\vecs k})\right),
~~~~P_{ij}(\hat{\vecs k})\equiv \delta_{ij}-{\hat k}_i{\hat k}_j,\label{proj}
\end{align}
with ${\hat k}_i \equiv k_i/|{\vecs k}|$. Here $h_{ij}({\vecs k},t)$ and $\Pi_{ij}({\vecs k},t)$ denote the Fourier transforms of $h_{ij}({\vecs x},t)$ and $\Pi_{ij}({\vecs x},t)$, respectively.
It was noted that it is sufficient to approximate $h_{ij}$ by $u_{ij}$ in the absence of spherical
symmetry and homogeneity~\cite{Q-ball:GW2}. The energy density of the GWs can be given by
\beqa
\rho_{\rm GW}&=&\frac{1}{32\pi G L^3}\int  d^3{\vecs x} {\dot u}_{ij}({\vecs x},t)
{\dot u}_{ij}({\vecs x},t)~,\nn\\
 &=&\frac{1}{32\pi G L^3}\int k^2 dk \int d\Omega~\Lambda_{ij,lm}(\hat{\vecs k}){\dot u}_{ij}({\vecs k},t){\dot u}_{lm}^*({\vecs k},t) \label{rhogw2},
\eeqa
where $V=L^3$ is the volume of the lattice.
%%%%and we have used the fact that the process of fragmentation is nonspherical and $|h_{ij}^{\rm TT}|\simeq |u_{ij}|$ in this situation.

GW reaches the maximal value at the last stage of the Q-balls formation, which can be estimated to be
\beqa
\rho_{GW}\simeq\f{M_{Pl}^2}{4}{\dot u}_{ij}{\dot u}_{ij}\simeq \f{\phi_Q^4}{36 M_{pl}^2}\beta_{\rm gr}^2~.
\eeqa
Here $\beta_{\rm gr}\equiv \dot{S}(t)$ is the fastest growing rate given by the evolution of the perturbation $\delta R \propto e^{S(t)-ik\cdot x}$ for $\Phi=R \exp(i\Theta)$. The fractional energy density parameter is
\beqa
\Omega_{GW}^*\simeq \f{\phi_Q^4}{108 M_{pl}^4 H_*^2}\beta_{\rm gr}^2.
\eeqa
 The fraction of the critical energy density stored in the GW today are
 \beqa
 \Omega_{\rm GW}^0&=&\Omega_{\rm
GW}^*\left(\frac{a_*}{a_0}\right)^4\left(\frac{H_*}{H_0}\right)^2\approx 1.67\tm 10^{-5}\(\f{100}{g_s^*}\)^{1/3}\Omega_{GW} h^{-2}.  \eeqa
The Hubble parameter $H_*$ at the $Q$-ball formation may be expressed as
$H_* =\beta_{\rm gr}/{\alpha}$, where $\alpha > 1$ is a numerical factor that represents the duration of the Q-balls formation.

%%, and $\alpha \simeq \CO(10)$ for the case of a mass term with negative $K$ \cite{Kusenko97}
%% and $\alpha \simeq \CO(1)$ for the case of a logarithmic term \cite{kk01}.
\subsection{Numerical Simulation}\label{3.3}
The non-linear property of the Q-balls formation and the subsequent evolution necessitates
a numerical simulation. We use the public code HLATTICE~\cite{HLattice} to simulate numerically the
fragmentation of AD fields and the formation of Q-balls with the potential of the form (\ref{xi:potential}). The evolution of the equation of motion for the homogeneous mode and the small perturbations $\delta R$
and $\delta\Theta$ can be solved numerically on a three dimensional cubic $N^3$ lattice with $N =128$.
In the HLATTICE package, the equations of motion are re-cast in a different form in order to make use
of more accurate, stable symplectic integrators.

%%%%%%%%%%%%%%%%%%%%%%%%%%Fig1%%%%%%%%%%%%%%%%%%
\begin{figure}[htb]
\begin{center}
\includegraphics[width=1.9in]{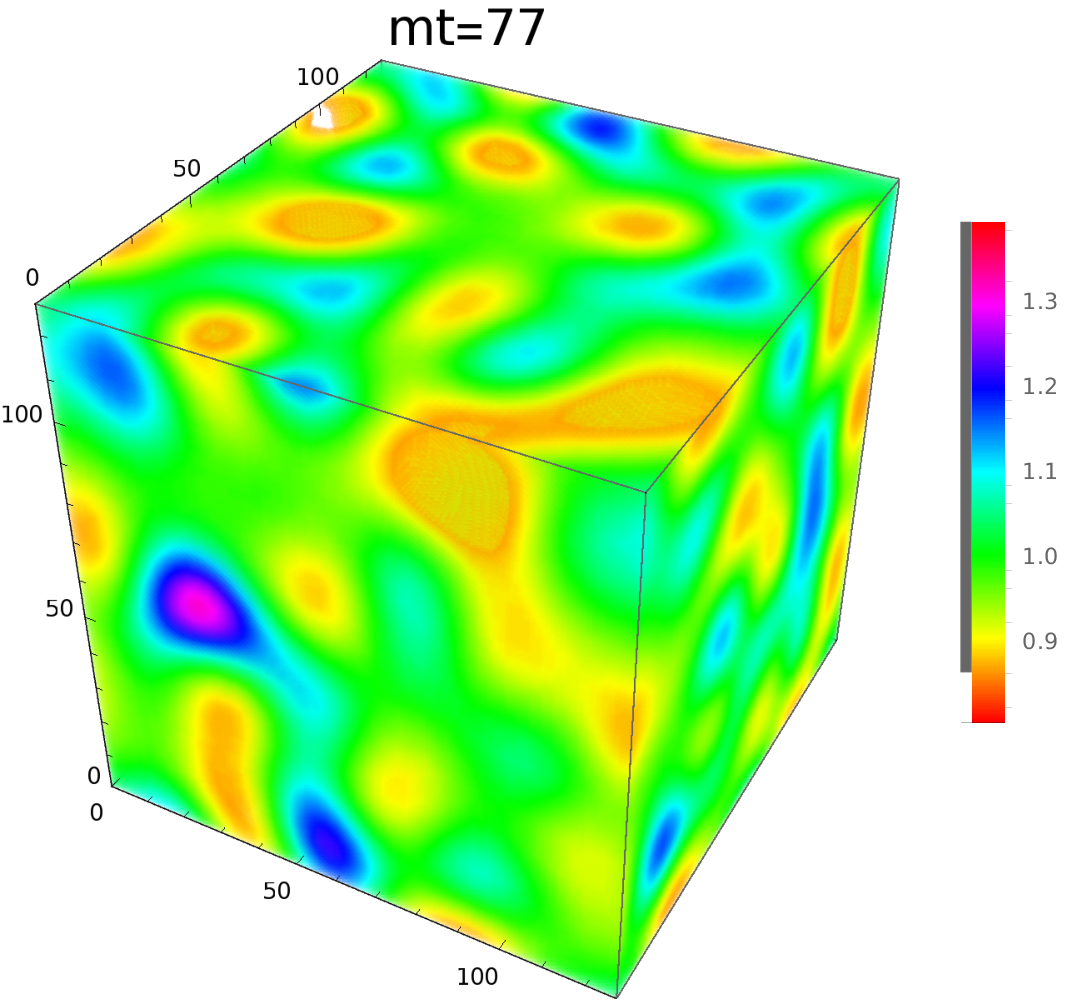}
\includegraphics[width=1.9in]{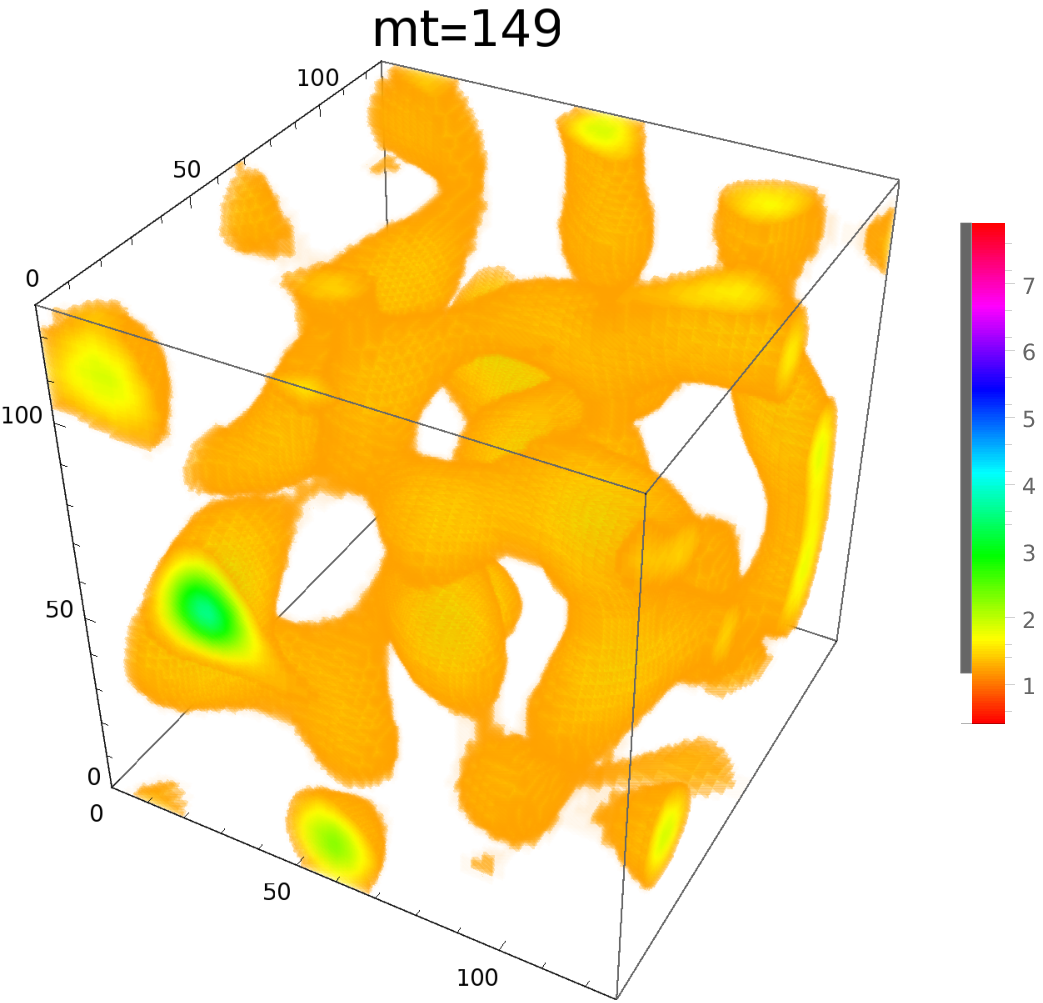}
\includegraphics[width=1.9in]{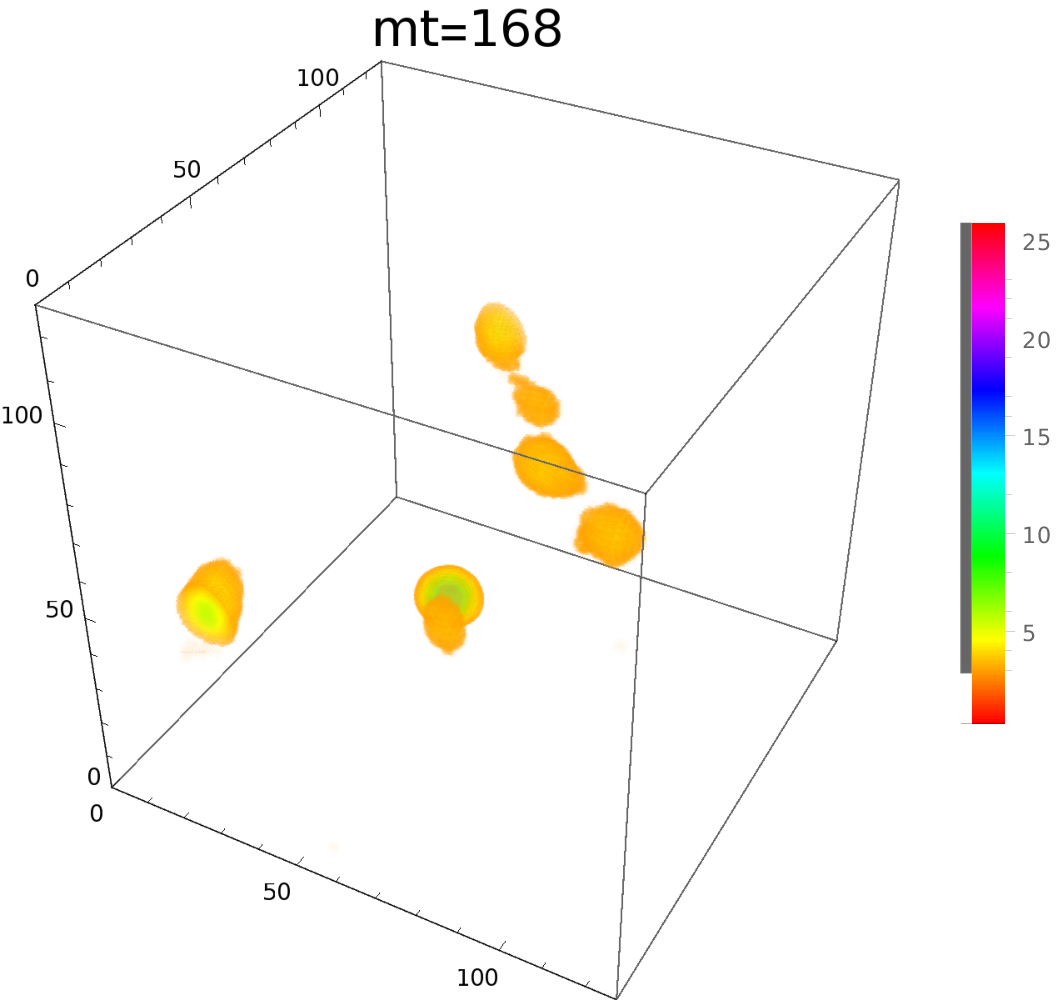}\\
\includegraphics[width=1.9in]{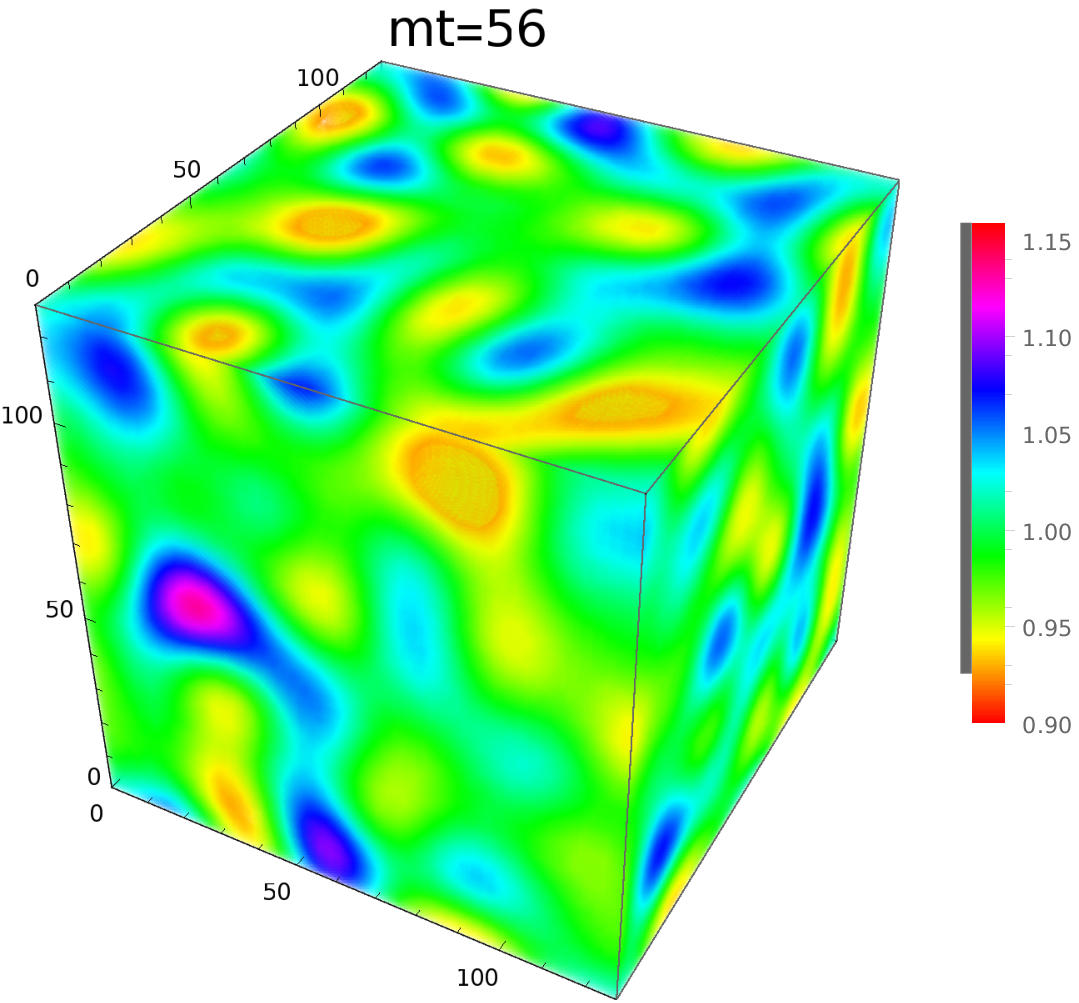}
\includegraphics[width=1.9in]{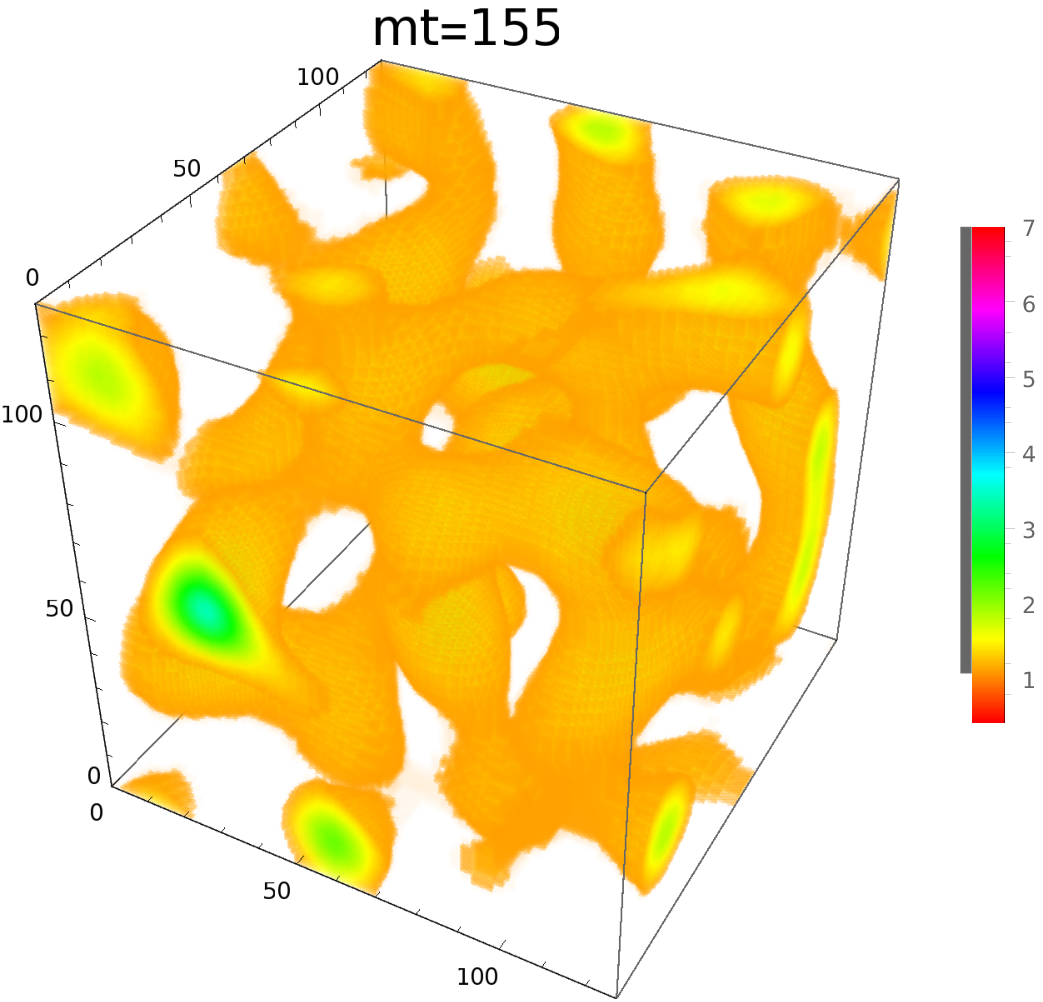}
\includegraphics[width=1.9in]{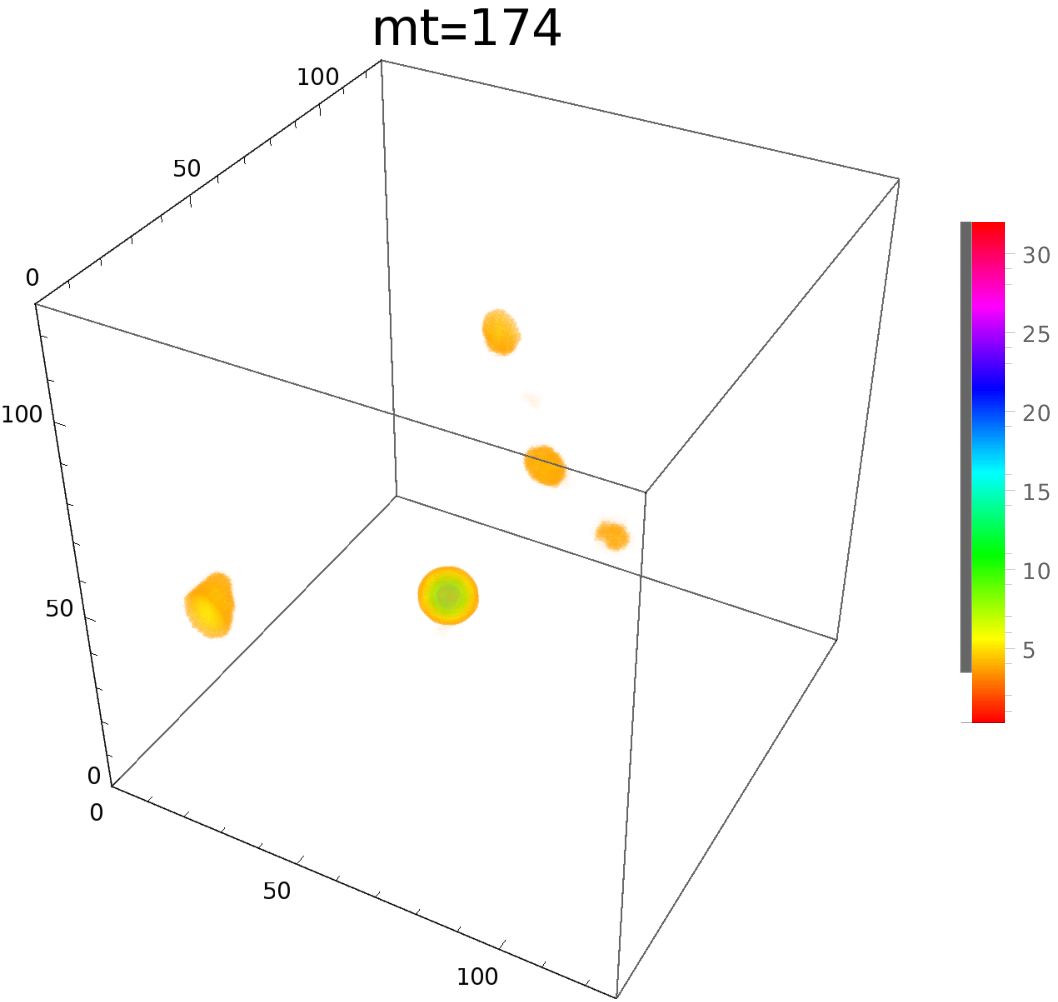}\\
\end{center}
\vspace{-.5cm}
\caption{The panels show the fragmentation of AD condensate and the formation of Q-balls in the case of
$K=0,\xi=10$ (upper panles) and $K=0.001,\xi=10$ (lower panels), respectively. The corresponding time scales are labeled in terms of
$mt$ with $m\equiv m_{3/2}$.
The input parameters are chosen as $m_{3/2}=100 {\rm TeV}$, $\phi_0=10^{17} {\rm GeV}$,~$\delta \phi/\phi_0=10^{-5}$,
$dx=50dt, LH=0.5$. The average energy density is normalized to 1.}
\label{fig1}
\end{figure}
%%%%%%%%%%%%%%%%%%%%%%%%%%%%%%%%%%%%%%%%%%%%%%%%%%%%%
The initial small fluctuations of the AD condensate come dominantly from the primordial quantum
fluctuations, which exited the horizon during inflation and re-entered the horizon afterwards.
From the inflationary cosmology, it can be expected to be $ |\delta\phi/\phi| \sim 10^{-5}$.
The final amplitude of the gravity waves is proved to be independent of the size of the initial
perturbation in the flat direction~\cite{Q-ball:GW2}.

%%%%%%%%%%%%%%%%%%%%%%%%%%Fig1%%%%%%%%%%%%%%%%%%
\begin{figure}[htb]
\begin{center}
\includegraphics[width=2.9in]{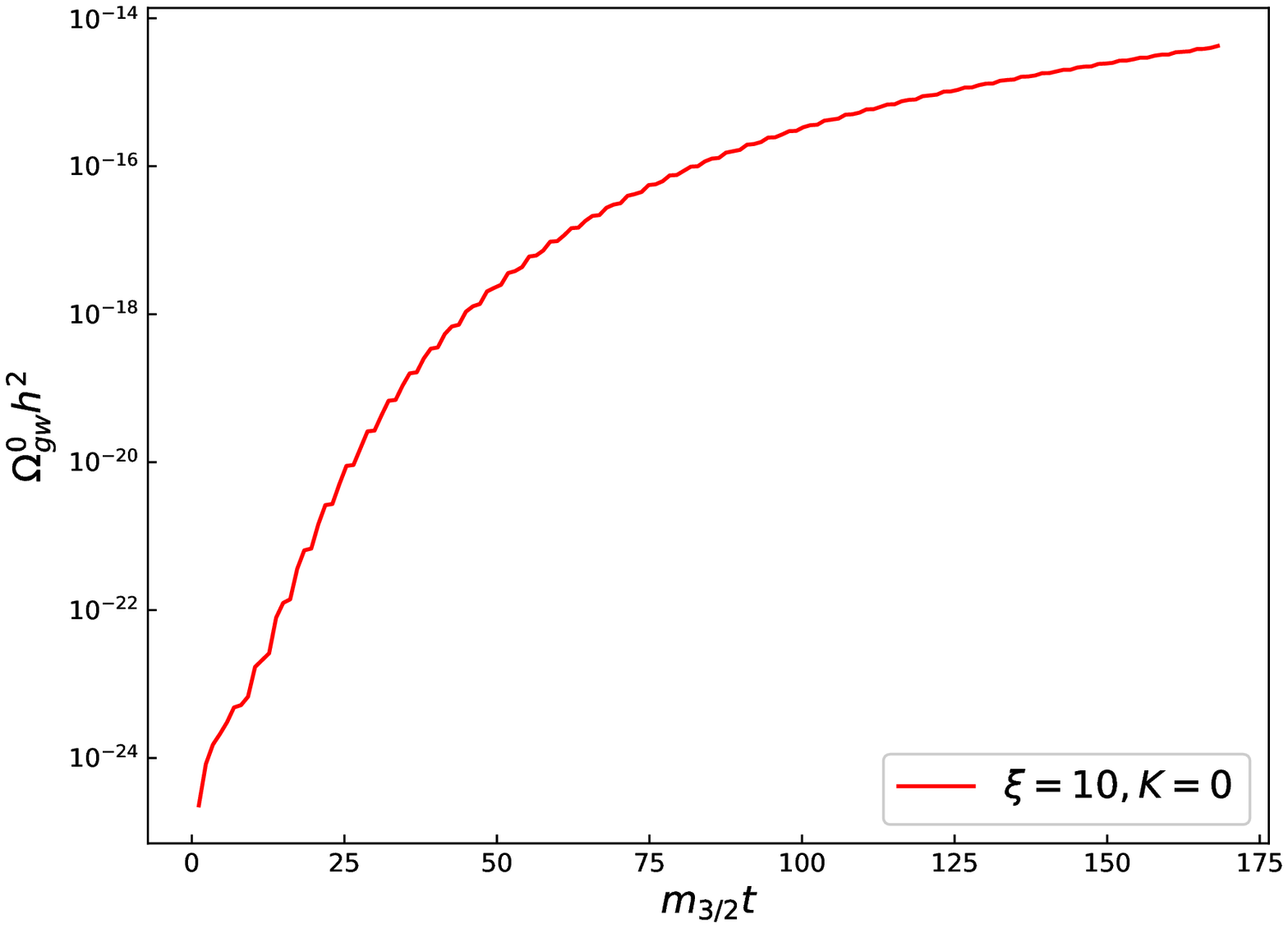}
\includegraphics[width=2.9in]{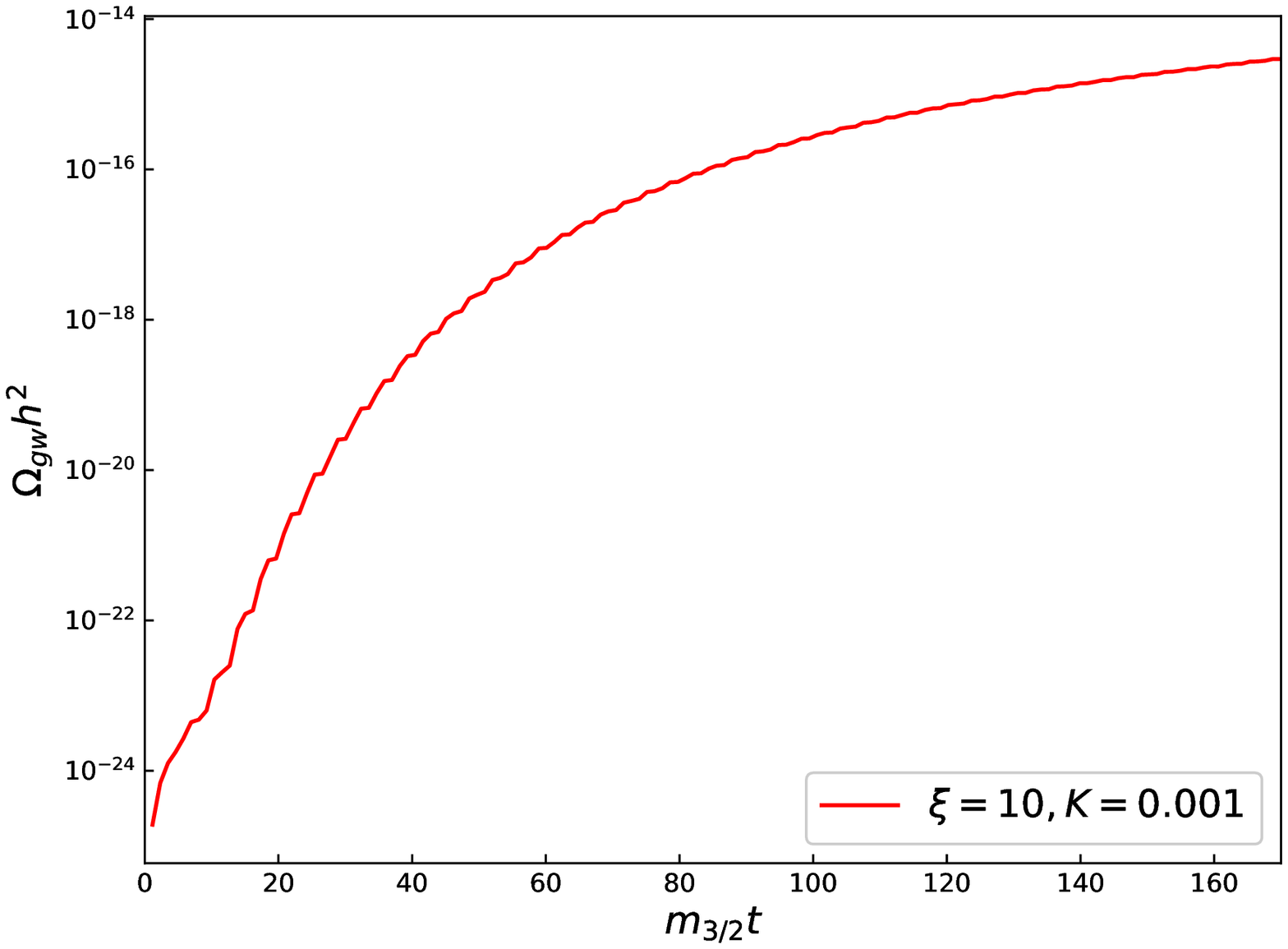}\\
\end{center}
\vspace{-.5cm}
\caption{ The panels show the temporal evolution of the present GW density parameter
$\Omega_{gw}^0h^2$ for the two benchmarks points
(in fig.1) with $K=0,\xi=10$ (left panel) and $K=0.001,\xi=10$ (right panel) , respectively. }
\label{fig2}
\end{figure}
%%%%%%%%%%%%%%%%%%%%%%%%%%%%%%%%%%%%%%%%%%%%%%%%%%%%%

We will discuss the Q-balls formation process and the emitted GWs with non-vanishing $\xi$, especially
the most interesting $K\geq 0,\xi>0$ case. The numerical results for benchmark points $K=0,\xi=10$ and
$K=0.001,\xi=10$ are shown in the panels of fig.\ref{fig1} and fig.\ref{fig2}.

In the simulation of the two benchmark points, the gravitino mass $m_{3/2}$ in the scalar potential
is chosen to be 100 TeV and the initial value of the homogeneous mode $\Phi_0$ is chosen to be $10^{17}{\rm  GeV}$,
which may correspond to the flat direction lifted by $n=9$. The AD condensate can fragment efficiently
into Q-balls when the fluctuations produced by linear parametric resonance become significant and the
dynamics become highly nonlinear. The process of the Q-balls formation can be seen in the panels
of fig.\ref{fig1}. From the panels, we can see the stages for AD fragmentation, the emerging of Q-balls
and the further evolution of Q-balls, respectively. It is clear that Q-balls can be formed successfully
in both cases.
%%%The maximum values of the energy density within the formed Q-balls are not large for positive $K$.
%%It can be understood that, unless the effects of non-minimal gravitational couplings are taken into account,
%%the Q-balls can not form with positive $K$. As such effects are not large, the maximum values of the energy density in the Q-balls should not be large either.

 The fragmentation process is not isotropic and  non-spherical motions of the condensate can
generate a quadrupole moment,
which will emit GWs during Q-balls formation.
The GWs are generated when the linear perturbation in the flat direction
condensate starts growing. The corresponding GW productions associated
with the fragmentation of
AD field in the case $K=0$ and $K>0$ can reveal the information of evolution.
The evolution of present GW density parameter
 as a function of the evolution
time $m_{3/2}t$ are shown in fig.\ref{fig2} for the benchmark points $K=0,\xi=10$ and
$K=0.001,\xi=10$, respectively.

%%%%%%%%%%%%%%%%%%%%%%%%%%Fig1%%%%%%%%%%%%%%%%%%
\begin{figure}[htb]
\begin{center}
\includegraphics[width=2.8in]{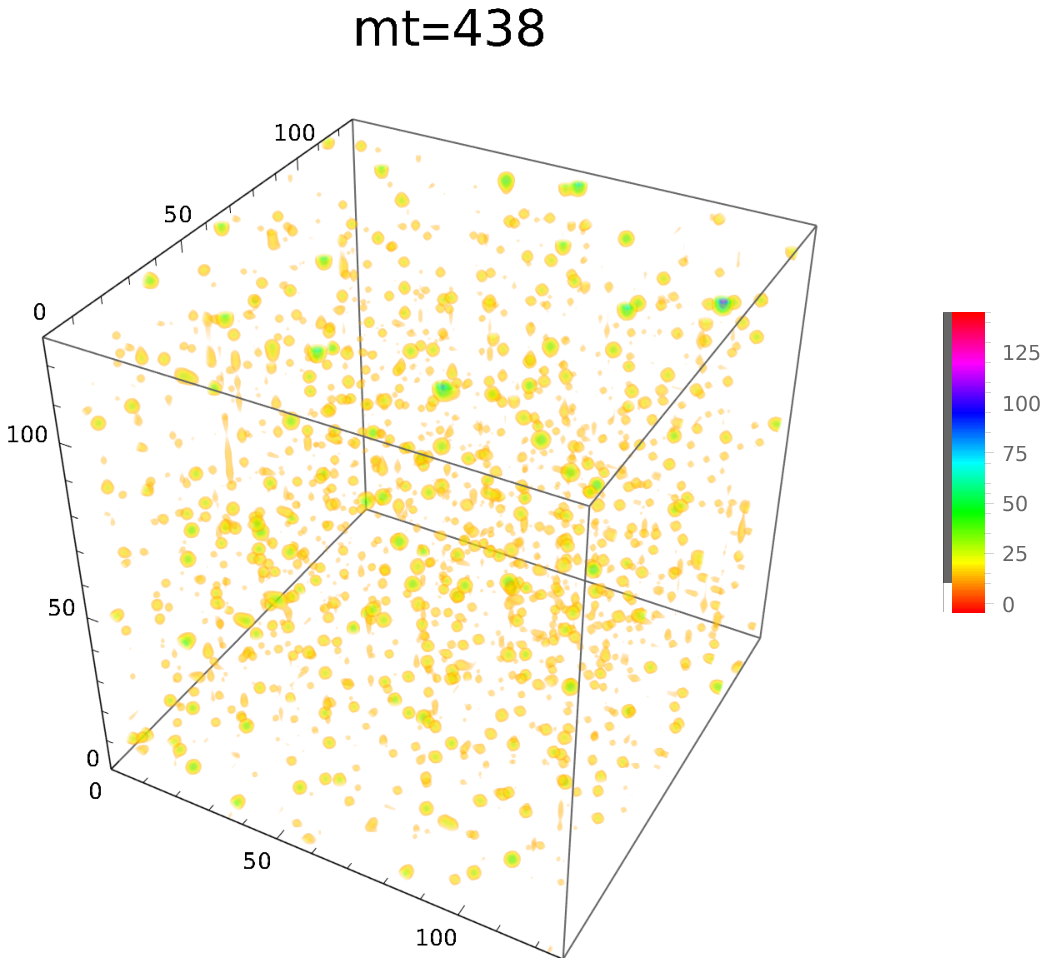}
\includegraphics[width=2.9in]{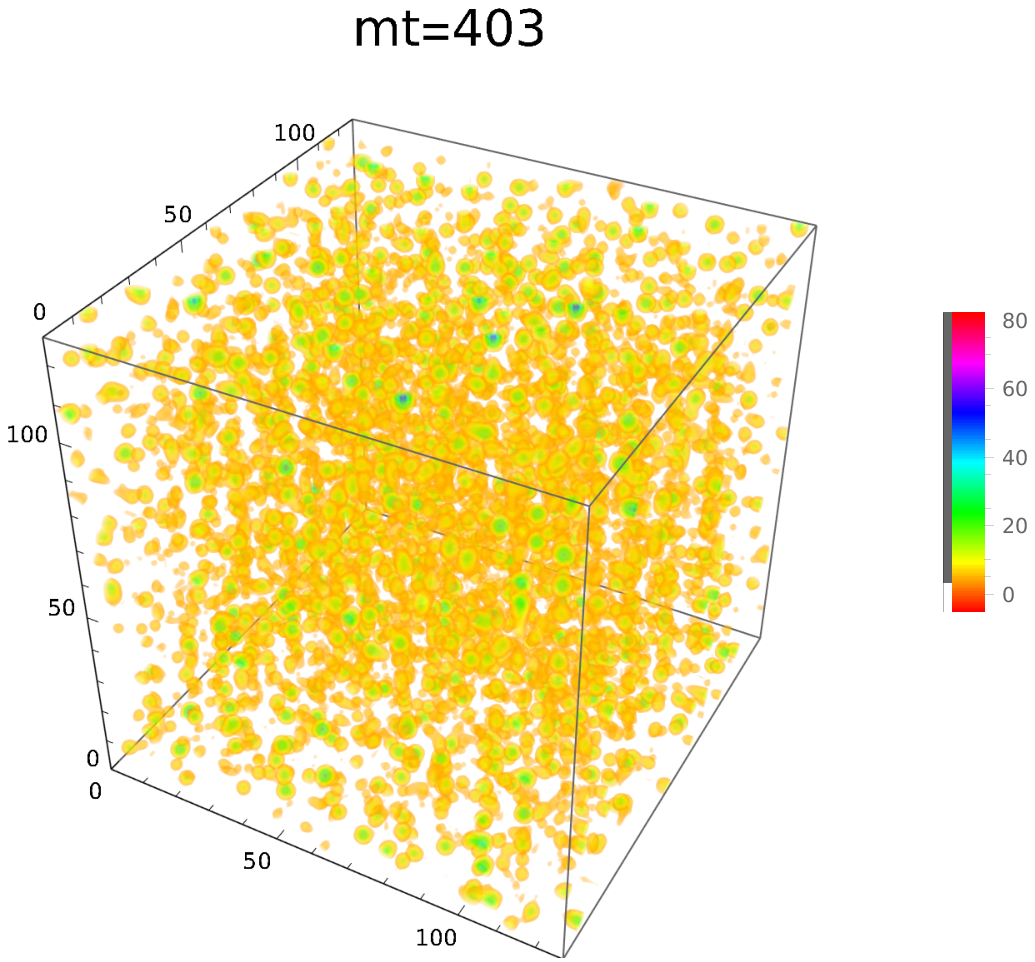}\\
\includegraphics[width=2.9in]{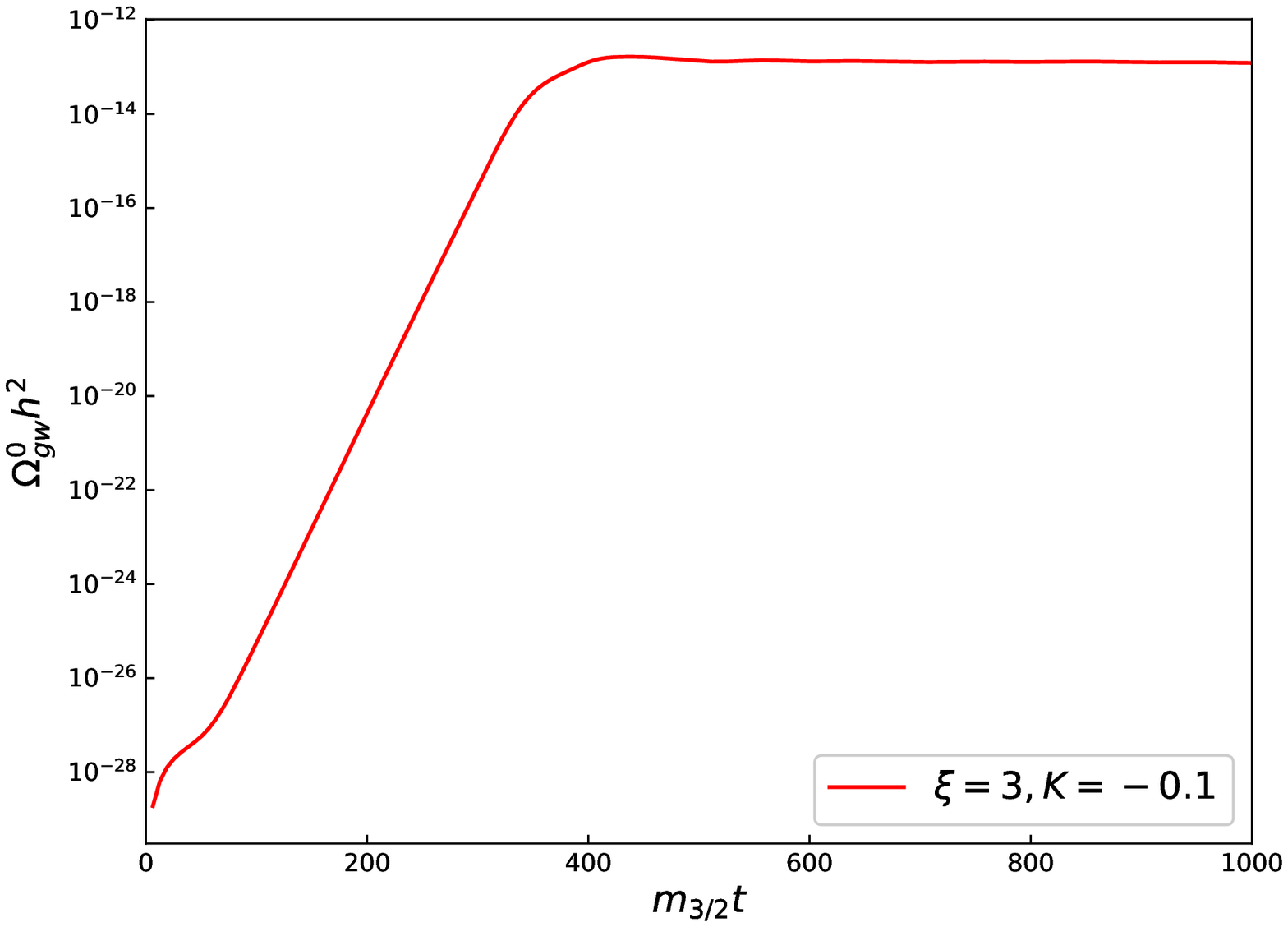}
\includegraphics[width=2.9in]{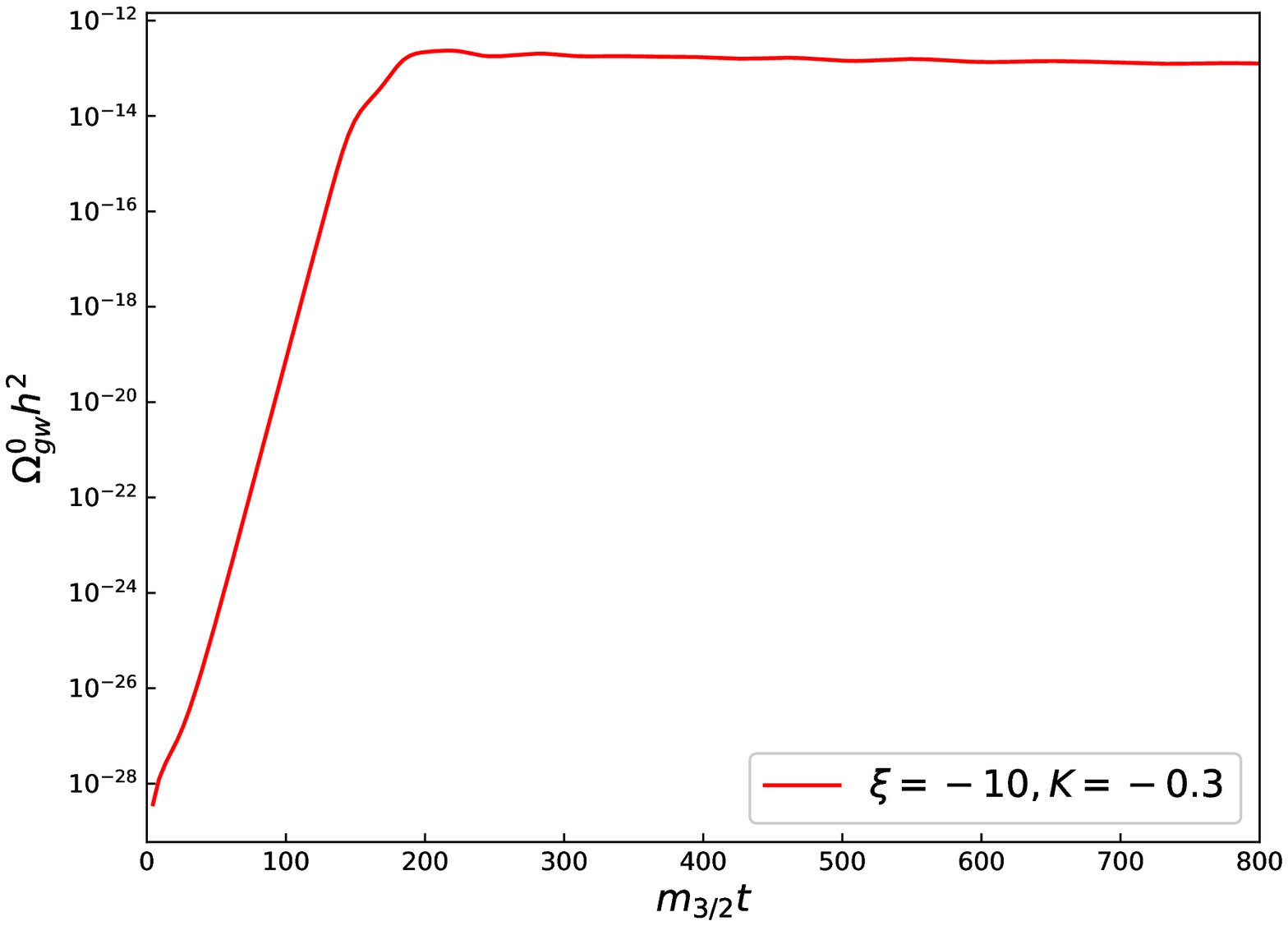}
\end{center}
\vspace{-.5cm}
\caption{Benchmark points with $K=-0.1,\xi=3$ (left panels) and $K=-0.3,\xi=-10$ (right panels).
The forms of the final-stage massive Q-balls
for both cases are shown in the upper panels.
The evolution of the GW density parameter
$\Omega_{gw}^0h^2$ with respect to $ m_{3/2} t$ are shown in the lower panels.
The input parameters
are chosen as $\phi_0=10^{16} {\rm GeV}$,~$\delta \phi/\phi_0=10^{-5}$,
$dx=50dt, LH=0.5$ with  $m_{3/2}=100 {\rm TeV}$. }
\label{fig3}
\end{figure}
%%%%%%%%%%%%%%%%%%%%%%%%%%%%%%%%%%%%%%%%%%%%%%%%%%%%%
  The form of the final-stage massive Q-balls and the evolution of the present GW density parameter
with negative $K$ and either sign of $\xi$ are also shown in fig.\ref{fig3}.
The setting of the input parameters
are shown in the caption of this figure.
 We show the benchmark points $K=-0.1, \xi=3$ and $K=-0.3, \xi=-10$ in the left and
  right panels of fig.\ref{fig3}, respectively.
Although the value $K=-0.3$ is relatively large in case of $\xi=0$ without
non-minimal gravitational couplings~\cite{Q-ball:GW2},
it is still acceptable with the choice of negative value $\xi$,
as discussed previously in subsection~\ref{3.1}.

%%%%%%%%%%%%%%%%%%%%%%%%%%Fig1%%%%%%%%%%%%%%%%%%
\begin{figure}[htb]
\begin{center}
\includegraphics[width=2.9in]{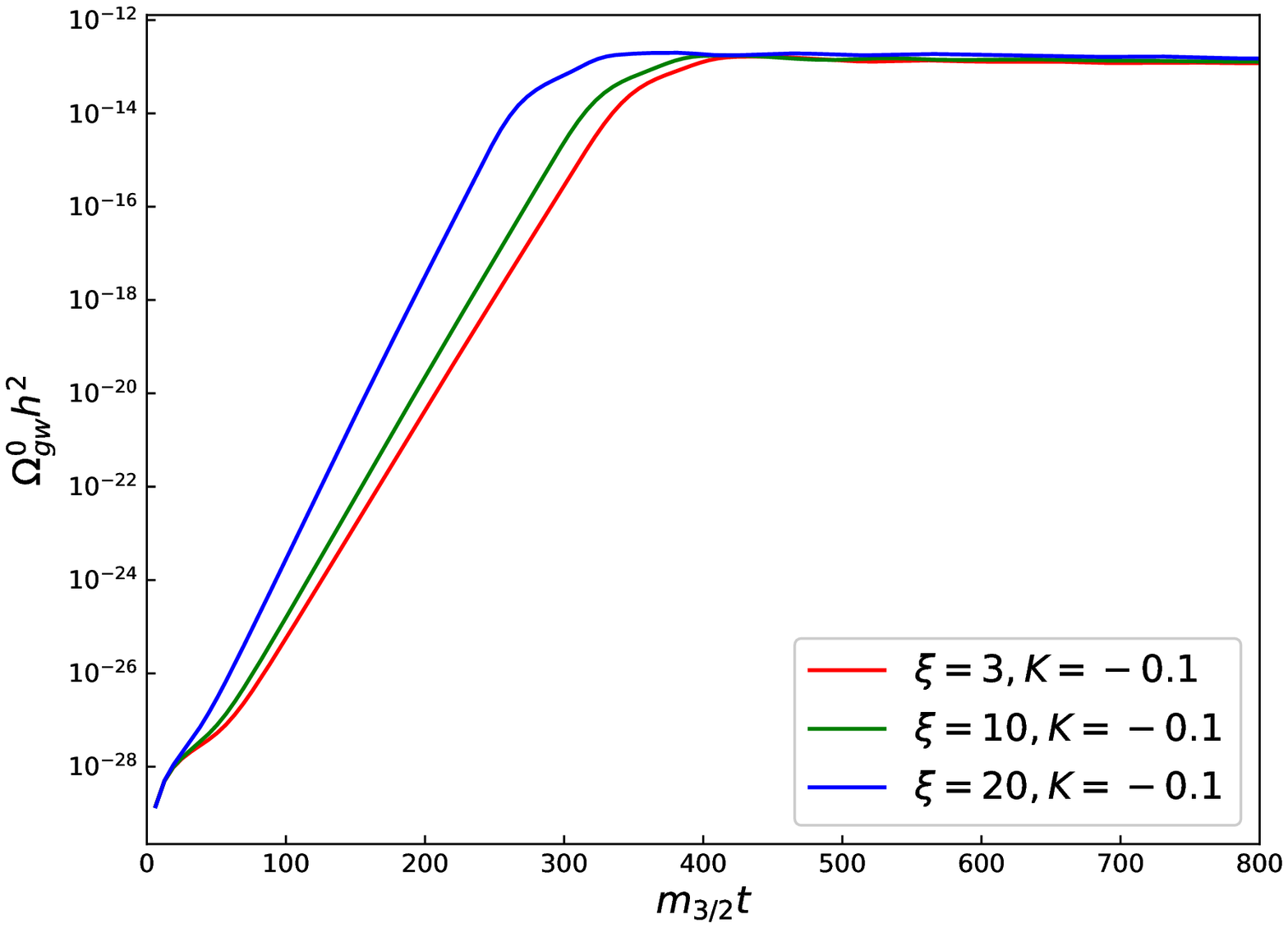}
\includegraphics[width=2.9in]{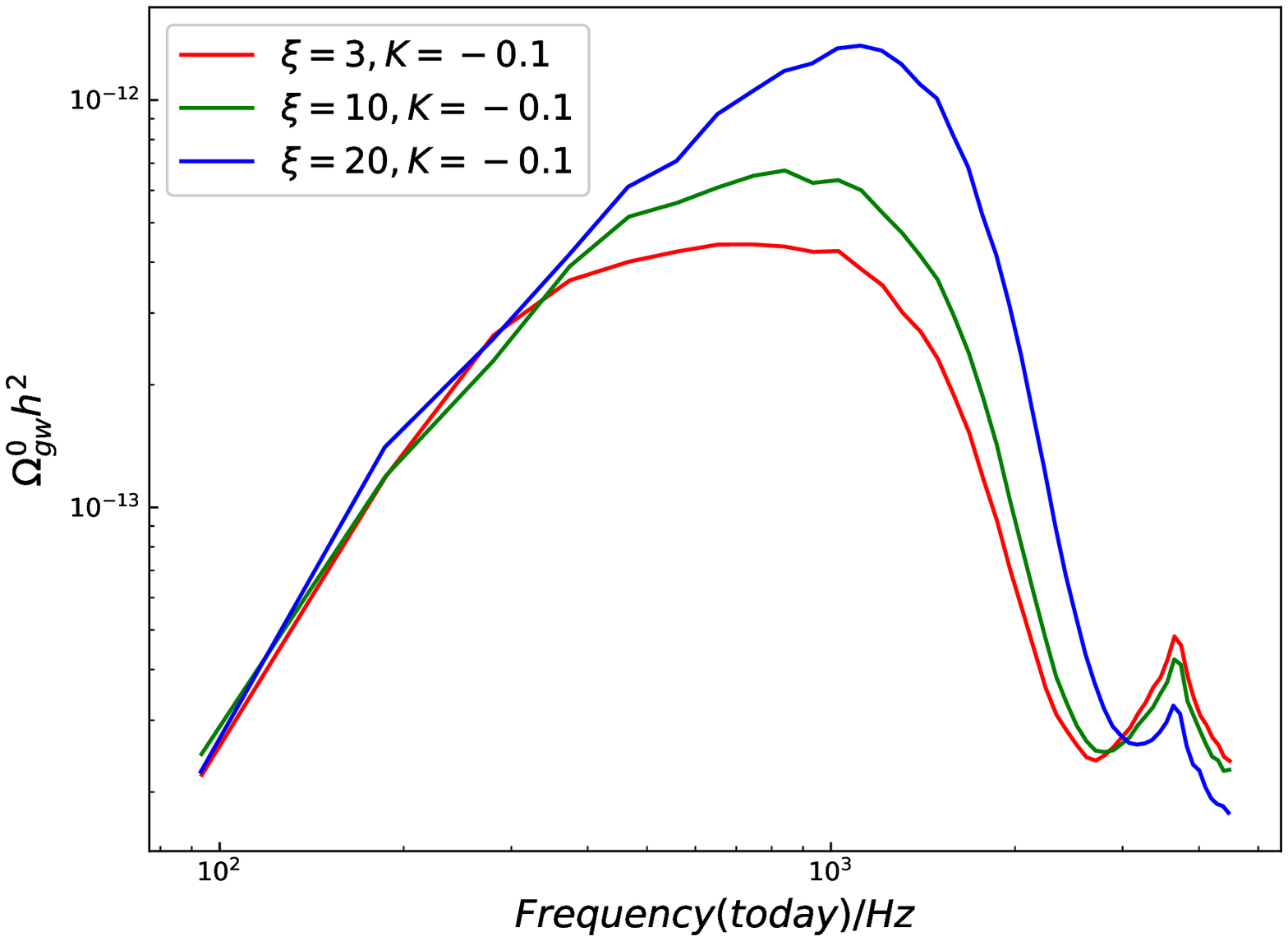}
\end{center}
\vspace{-.5cm}
\caption{Same input parameters as fig.3. The evolution of the present GW density parameter
$\Omega_{gw}^0h^2$ and its dependence
on the choices of $\xi$ are shown in the left panel.
The frequency bands of the GWs from Q-ball formation at the frequency ranging from
$100$ Hz to $10^3$ Hz in each case are shown in the right panel.}
\label{fig4}
\end{figure}
%%%%%%%%%%%%%%%%%%%%%%%%%%%%%%%%%%%%%%%%%%%%%%%%%%%%%
The GW power spectrum for $K=-0.1$ with various value of $\xi$ are shown in fig.\ref{fig4}.
We can see that the final amplitude of the present GW density parameter
$\Omega_{gw}^0h^2$ during the evolution is insensitive to the value of $\xi$.
Different choices of $\xi$ lead to different growth rate of the GWs.
The peak frequency $f_{P}$ of $\Omega_{gw}^0h^2$
is also insensitive to the choices of $\xi$
while the peak values of $\Omega_{gw}^0h^2$ depend
on the choices of $\xi$. Larger $\xi$ will lead to larger peak value of $\Omega_{gw}^0h^2$.
The peak frequency for GW power spectrum lies around a few KHz with our choice of $m_{3/2}=100 {\rm TeV}$.
  Therefore, the stochastic GW backgrounds can not be observed by the current and upcoming interferometer based GW experiments.
  It was shown in~\cite{1501.01217} that the peak position of the GW power spectrum depends on the value of $m_{3/2}$.
  Larger values of $m_{3/2}$ lead to higher GW frequencies. Choosing a lower value of $m_{3/2}$ can shift the spectrum to
  lower frequency. However, we find that such a small-shifted GW power spectrum can still not be detected by the upcoming
  interferometer based GW experiments.  As the gravitino mass is given by $m_{3/2}=F/\sqrt{3}M_P$, low SUSY breaking scale
  $F$ may cause low GW frequencies. If such stochastic GW signal are detected,
it may give interesting information on the value of SUSY breaking scale.

We should note that our calculations ignore reheating and thermalization due to the AD condensate. Once the decay of the condensate is taken into account, the amplitude of the GWs  will be attenuated. Such topics will be discussed in our subsequent studies. It is known that finite-temperature corrections to the flat direction potential can act as a source of SUSY breaking, which could have important effects on the dynamics of flat directions. Thermal corrections, whose exact effect depend on the temperature of the thermal bath $T$ and the nature of the AD condensate, can be important if the inflaton decays dominantly into the visible sector fields and produces MSSM degrees of freedom. In general, a precise calculation for the baryon asymmetry via the AD mechanism and GW productions should take them into account~\cite{Allahverdi:2012}.
%%%Since the AD condensate is made up of MSSM scalars, the finite-temperature corrections would only involve those degrees of freedom that are coupled, directly or via loops, to these scalars.

 Simple treatment of thermal effects are based on the existence of a thermal plasma from inflaton decay at arbitrarily early times, which implicitly assumes that particles produced from decay of the inflaton immediately reach thermal equilibrium. In fact, assignment of a temperature $T$ to the reheat plasma is only justified after full thermal equilibrium is achieved. However, reheating and thermalization after inflation can be a very complicated process involving various perturbative and non-perturbative phenomena. For example, the helicity $\pm 3/2$ gravitino can be produced non-perturbatively from vacuum fluctuations after inflation~\cite{Maroto:1999ch}. Particles produced from inflaton decay typically have a non-thermal distribution, and the time scale of their equilibrium is model dependent.
% The coherent oscillations of the inflaton can create particles non-perturbatively and such preheating is particularly efficient when the final products are bosonic degrees of freedom.
In a non-SUSY case, the inflaton usually decays via preheating unless its couplings
to other fields are very small. The inflaton decay products thermalize very quickly because of the efficiency of interactions mediated by the massless gauge bosons of the SM. Therefore, the reheat temperature is mainly governed by the inflaton decay width. However, preheating is unlikely within supersymmetry because flat directions in the scalar potential are generically displaced towards a large VEV in the early Universe, which induce supersymmetry preserving masses to the inflaton decay products and consequently prohibit non-perturbative inflaton decay into MSSM fields~\cite{Allahverdi:2006pfc}.  Besies, the process of thermalization within SUSY is in general very slow~\cite{Allahverdi:2005fq,Allahverdi:2005mz} because the VEV of the AD condensate can induces a large mass to the gauge fields via Higgs mechanism, thereby suppressing the rate of processes relevant for thermalization. If the entire SM gauge group is broken, thermalization can be delayed substantially. A full thermal equilibrium is generically established much later on when the VEV of the flat direction has substantially decreased. Therefore, Universe loiters in a phase of a quasi-thermal equilibrium after the decay of the inflaton, resulting in a very low reheat temperature, perhaps as low as {\cal O}(TeV). The final reheat temperature depends on a thermalization time scale instead of the decay width of the inflaton.
%%Since the AD condensate is made up of fields that carry gauge quantum numbers, its VEV
%%spontaneously breaks gauge symmetries. This induces a large mass ¡«g|¦Õ| to the gauge fields via
%% the Higgs mechanism, thereby suppressing the rate of processes relevant for thermalization.
%%If the entire SM gauge group is broken, thermalization can be delayed substantiallyIn fact, flat directions that are lifted by superpotential terms of order n ? 6, see equation (3), can result in reheat temperatures well below the naive estimate TR ¡« (?dMP)1/2, perhaps as low as O(TeV)

\section{Conclusions}\label{sec-4}
We propose to introduce non-minimal couplings of Affleck-Dine (AD) field to gravity by adding the
coupling of AD field to the Ricci scalar curvature.
A possible realization to generate the $c_0|\Phi|^2 R$ type coupling terms with a general $c_0$ in Jordan frame supergravity after SUSY breaking is given.
The impacts of such non-minimal gravitational couplings for AD field is shown, especially
on the Q-balls formation and the associated gravitational wave (GW) productions.
New form of scalar potential for AD field in the Einstein frame is obtained.
By numerical simulations, we find that, with non-minimal gravitational coupling to AD field,
Q-balls can successfully form even with the choice of non-negative $K$ parameter for $\xi>0$.
The associated GW productions as well as their dependences on the $\xi$ parameter are also discussed.\addcontentsline{toc}{section}{Acknowledgments}
\acknowledgments
We are very grateful to the referee for helpful discussions and very useful suggestions. We acknowledge Ligong Bian for discussions. This work was supported by the
Natural Science Foundation of China under grant numbers 12075213,11675147; by the Key Research Project
of Henan Education Department for colleges and universities under grant number 21A140025.

\end{document}